\documentclass[aps,amsmath,amssymb,prb,twocolumn,showpacs,superscriptaddress]{revtex4-1}
\usepackage{bm}
\usepackage{epsfig}
\usepackage[usenames]{color}
\usepackage{graphicx}
\usepackage[hypertex]{hyperref}

\DeclareMathOperator{\acosh}{acosh}

\newcommand{\nn}{\nonumber}
\newcommand{\inel}{{\rm in}}
\newcommand{\ph}{{\rm ph}}
\newcommand{\T}{{\mathfrak{T}}}
\newcommand{\dt}{{\tau}}
\newcommand{\V}{{\mathcal{V}}}

\newcommand{\PS}{{{\cal P}_{\rm s}}}
\newcommand{\PSd}{{\mathfrak{P}_{\rm s}}}

\newcommand{\half}{1/2} 

\renewcommand{\paragraph}[1]{\textit{#1.---} } 

\begin{document}

\title{Heating Effects in a Chain of Quantum Dots}

\author{A.~Glatz}
\affiliation{Materials Science Division, Argonne National Laboratory, Argonne, Illinois 60439, USA}

\author{I.~S.~Beloborodov}
\affiliation{Department of Physics and Astronomy, California State University Northridge, Northridge, CA 91330, USA}

\author{N.~M.~Chtchelkatchev}
\affiliation{Materials Science Division, Argonne National Laboratory, Argonne, Illinois 60439, USA}
\affiliation{Institute for High Pressure Physics, Russian Academy of Sciences, Troitsk 142190, Moscow region, Russia}
\affiliation{L.D.\ Landau Institute for Theoretical Physics, Russian Academy of Sciences, 117940 Moscow, Russia}

\author{V.~M.~Vinokur}
\affiliation{Materials Science Division, Argonne National Laboratory, Argonne, Illinois 60439, USA}

\date{\today}

\begin{abstract}
We study heating effects in a chain of weakly coupled grains due to electron-hole pair creation.
The main mechanism for the latter at low temperatures is due to inelastic electron cotunneling processes in the array.
We develop a quantitative kinetic theory for these systems and calculate the array temperature profile as a function of grain parameters, bias voltage or current, and time and show that for nanoscale size grains the heating effects are pronounced and easily measurable in experiments. In the low- and high-voltage limits we solve the stationary heat-flux equation analytically. We demonstrate the over-heating hysteresis in the large-current or voltage regimes. In addition we consider the influence of a substrate on the system which acts as a heat sink. We show that nano dot chains can be used as highly sensitive thermometers over a broad range of temperatures.
\end{abstract}

\maketitle

\section{Introduction}

Great efforts in contemporary materials science research focus on the understanding of thermal properties of nanoscale devices~\cite{Glatz09b,RMP06,Pekola07,BeloborodovEPL05,glatz+prb09b,Saira07}.
The electron transport through arrays of nanogranular systems is well understood~\cite{Beloborodov07,BeloborodovPRL,Efetov02,Fogler,Beloborodov05,CVB_PRL,Rodionov} at small currents when the charge carriers have enough time to equilibrate their temperature with the temperature of the phonon bath. It is known that most of the transport characteristics of the system, e.g., the resistance, are in fact very sensitive to the temperature of the electronic excitations in the granules. It is much less known about {\it self-heating} effects in these systems when the current is not very small and the charge flow makes the internal temperature different (higher) than the temperature of the phonon bath. Understanding of the heating effects on properties of nanodevices is especially important for practical applications. In addition, recent experimental research~\cite{Heath08,Shi03,Hoffmann07} has focused on the application of the thermal properties of low-dimensional devices for efficient power generation. Granular thermoelectric materials have the advantage that one can control the system parameters and therefore the device properties. Due to the high sensitivity of the resistance of granular materials on temperature these materials are good candidates for broad band temperature sensors that can operate with high resolution from cryogenic up to room temperatures. This defines an urgent quest for a quantitative description of heating effects on properties of nanodevices.

\begin{figure}[tbh]
\includegraphics[width=0.8\columnwidth]{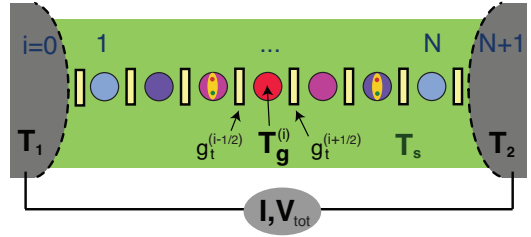}
\caption{(Color online) Sketch of a weakly coupled nano dot chain of length $N$ connected to two leads, with tunneling conductances $g^{(i\pm\half)}_t \ll 1$ (where $i\pm\half$ denotes the tunneling barriers right/left of grain $i$). The leads have the grain indices $0$ and $N+1$. The lead temperatures $T_1$ and $T_2$ and substrate temperature $T_s$ are fixed.  The grain temperature $T^{(i)}_g$ depends on these temperatures and on the applied current or voltage. The main mechanism of heating in the grain due to inelastic cotunneling processes is depicted by the electron-hole pairs inside the grain which have a typical energy of $T^{(i)}_g$. The colors of the grains show a typical distribution of grain temperatures throughout the system: the coldest grains are close to the leads and the hottest in the middle of the chain.}
\label{fig.model}
\end{figure}

In this paper we investigate the effect of inelastic cotunneling on heating of weakly coupled nano dot chain of length $N$ coupled to two leads (see Fig.~\ref{fig.model}). We calculate the grain temperature profile as a function of grain parameters, bias current, and time for different system setups of one-dimensional chains, which may or may not be coupled to a substrate through electron-phonon collisions.

Each grain is characterized by two energy scales: (i) the mean energy level spacing $\delta$ and (ii) the charging energy $E_c = e^2(4\pi\kappa\epsilon_0 a)^{-1}$, where $\kappa$ is the relative static permittivity of the grain material. We concentrate on the case of metallic grains which are most commonly encountered in applications satisfying the inequality $E_c \gg \delta$.

In addition to the above energy scales the system under consideration is characterized by the tunneling conductances $g^{(i)}_t$. In this paper we concentrate on the regime of Coulomb blockade, meaning that the tunneling conductances are much smaller than the quantum conductance, $g^{(i)}_t \ll 1$.  Our considerations are valid for temperatures $ \delta < T < E_c$.

\section{Mechanism for grain heating}

In the considered low temperature region, the only transport mechanism contributing to grain heating is due to inelastic cotunneling. Cotunneling allows simultaneous charge transport through several junctions by means of cooperative electron motion. Single cotunneling (Fig.~\ref{fig.model}), introduced in Ref.~[\onlinecite{Averin}], see also the review~\cite{Grabert_book}, provides a conduction channel at low applied biases, where otherwise the Coulomb blockade arising from electron-electron repulsion would suppress the current flow. The essence of a cotunneling process is that an electron tunnels via virtual states in intermediate granules thus bypassing the huge Coulomb barrier. This can be visualized as coherent superposition of two events: tunneling of the electron into a granule and the simultaneous escape of another electron from the same granule. There are two distinct mechanisms of cotunneling processes, elastic- and inelastic cotunneling. Elastic cotunneling means that the electron that leaves the dot has the same energy as the incoming one. In the event of inelastic cotunneling, the electron coming out of the dot has a different energy than the entering electron. This energy difference is absorbed by an electron-hole excitation in the dot, which is left behind in the course of the inelastic cotunneling process, depicted in Fig.~\ref{fig.model}.
Below we concentrate on inelastic cotunneling because only this transport mechanism contributes to heating effects.
In particular, elastic cotunneling and sequential tunneling do not create electron-hole pairs in the grain.

For the quantitative description of the non-equilibrium effects leading to the heating of the chain, we consider (i) the heating due to electric resistance, (ii) the heat removal due to electron and phonon thermal conductivities, and (iii) in some cases the electron-phonon coupling of the array to a substrate. The electric conductivity and the electronic part of the thermal conductivity are governed by inelastic cotunneling processes.
Using all these ingredients we can write down the kinetic equation for the grain temperatures in the following section, describing the time evolution and final temperature profile over the chain.

\section{Kinetic heat equation}

For the temperatures of each grain $i$ in the array we can write down the following kinetic equation
\begin{equation}
\label{eq.model}
c^{(i)}_v \frac{\partial T^{(i)}_g}{\partial t} =\sigma_{\inel}^{(i)}\left(\frac{V^{(i)}}{a}\right)^2
+ \hat\nabla\left(\kappa^{(i)}\hat\nabla T^{(i)}_g\right)-\PS^{(i)},
\end{equation}
where $c^{(i)}_v = (1/3) k_B^2 T^{(i)}_g \nu$ is the heat capacitance of each grain with $\nu=1/(a^3\delta)$ being
the density of states in the grain;
$\partial T^{(i)}_g/\partial t$ describes the change of the grain
temperatures $T^{(i)}_g$ in time $t$.
$\sigma_{\inel}^{(i)}$ is the cotunneling conductance (defined below) for grain $i$, $V^{(i)}$ being the  voltage drop and $a$ the grain size; $\kappa^{(i)}$ is the thermal conductivity, with $\kappa^{(i)}=\kappa^{(i)}_\inel+\kappa^{(i)}_\ph$ where $\kappa^{(i)}_\inel$ is the electron and $\kappa^{(i)}_\ph$ the phonon contribution to the thermal conductivity:
for a single metallic grain weakly
coupled to its neighbors or the leads these given by the following expressions~\cite{Glazman,Averin,Tripathi06,Beloborodov07,Glatz09}
\begin{subequations}\label{eqs.conduct}
\begin{eqnarray}
\sigma^{(i)}_\inel &=& 2e^2 a^{-1} \left(g^{(i)}_t\right)^2 \frac{\left(k_B T^{(i)}_g\right)^2 + \left(eV^{(i)}\right)^2}{\hbar E_c^2} \label{eq.sigma}, \\
\kappa^{(i)}_\inel &=& \gamma_e a^{-1} \left(g^{(i)}_t\right)^2 k_B^2 T^{(i)}_g \frac{\left(k_B T^{(i)}_g\right)^2 + \left(eV^{(i)}\right)^2}{\hbar E_c^2}\label{eq.kappae},\\
\kappa^{(i)}_\ph &=& \gamma_\ph (\hbar l_\ph)^{-1} k_B^2 T^{(i)}_g \left(\frac{T^{(i)}_g}{\Theta_D}\right)^2.\label{eq.kappaph}
\end{eqnarray}
\end{subequations}
In Eq.~(\ref{eq.kappae}) and (\ref{eq.kappaph}) $\gamma_{e/\ph}$ are numerical coefficients ($\gamma_e=32\pi^3/15$ and $\gamma_\ph=8\pi^2/15$, see [\onlinecite{Glatz09}]), $l_\ph$ the phonon mean free path $l_\ph\sim a$, and $\Theta_D$ the Debye temperature. Notice, that the temperature entering Eqs.~(\ref{eqs.conduct}) is the grain temperature, $T^{(i)}_g$, which determines the number of typically involved internal energy states of the grain. The tunneling conductances $g^{(i)}_t$ are to be understood as the averaged quantities $\sqrt{g^{(i-\half)}_tg^{(i+\half)}_t}$, with $i\pm\half$ denoting the tunneling barriers right/left of grain $i$.

The symbol $\hat\nabla$ in Eq.~(\ref{eq.model}) refers to the discrete gradient along the chain.
The last term in the r.h.s. of Eq.~(\ref{eq.model}), $\PS^{(i)}$ describes the heat dissipation from grain $i$ due to electron-phonon coupling of the array to a substrate and is given by~\cite{Pekola07}
\begin{equation}\label{eq.PS}
\PS^{(i)}= \Sigma_s\left[\left(T^{(i)}_g\right)^5 - T_s^5\right]\,,
\end{equation}
with $T_s$ being the substrate temperature and $\Sigma_s$ the electron-phonon coupling constant, which is of the order $10^4$W/(m$^3$K$^5$). This electron-phonon term is written for the {\it clean} case meaning that the phonon mean free path $\lambda_{ph}$ is smaller than the electron length, $\lambda_{ph} < l_{el}$,~\footnote{For quantum dot arrays the phonon length $\lambda_{ph}$ is of the order of grain size, $\lambda_{ph}\sim a$.}. Here, we notice that the applied current density $j$ does not influence the substrate term.

Equation~(\ref{eq.model}) has a transparent physical meaning:
The energy of the electron-hole pair is released as heat, see also Fig.~\ref{fig.model} whereas the thermal conduction to the leads removes the heat from the grain array until a steady state is reached.
The typical energy of electron-hole pairs is of order of the grain temperatures, $T^{(i)}_g$, and included in the expressions for $\sigma_{\inel}^{(i)}$ and $\kappa_{\inel}^{(i)}$.
The leads are assumed to be a heat sink and stay at ambient temperature $T$ since they are much larger than the grains.
The first term on the r.~h.~s. of Eq.~(\ref{eq.model}) describes the energy released in the grain from an electron-hole pair times the square of the discrete gradient of voltage ($V^{(i)}/a$).
The latter ensures that the result is invariant under a sign change of the voltage $V^{(i)}$. The second term $\hat\nabla\left(\kappa^{(i)}\hat\nabla T^{(i)}_g\right)$ describes the heat-flux out of grain $i$.
Our considerations are valid as long as the applied voltage does not break down the Coulomb blockade, i.e. when $eV^{(i)}<E_c$.

In typical experiments the current is kept constant and Eq.~(\ref{eq.model}) can be written as
\begin{equation}\label{eq.model2}
\hat\nabla\left(\kappa^{(i)}\hat\nabla T^{(i)}_g\right)=c^{(i)}_v \frac{\partial T^{(i)}_g}{\partial t}-\frac{\left(j^{(i)}\right)^2}{\sigma_{\inel}^{(i)}}+\PS^{(i)}\,,
\end{equation}
where we introduced the current as $j^{(i)}=\sigma_{\inel}^{(i)}\frac{V^{(i)}}{a}$ which is constant through the whole chain. Although this equation is equivalent to Eq.~(\ref{eq.model}) it has a different physical interpretation, since it describes the heat flux through the granular array. In the following we consider the case of constant current.

\subsection{Kinetic heat equation in dimensionless units}

It is convenient to rewrite the kinetic Eq.~(\ref{eq.model2}) in dimensionless units. Temperatures and voltages are measured in units of the first lead temperature $T_1$ and the time scale is in units of mean energy level spacing $\delta$, i.e.
\begin{eqnarray}
\label{dimensionless1}
\T^{(i)}_g=T^{(i)}_g/T_1&,&  \hspace{0.3cm} \T_2=T_2/T_1, \hspace{0.3cm} \T_s=T_s/T_1,\nn\\
 \V^{(i)}=eV^{(i)}/(k_B T_1)&,& \hspace{0.3cm} \dt=t\delta/\hbar,
\end{eqnarray}
where $\T^{(i)}_g$, $\T_2$, $\T_s$ are the grain, second lead, and substrate dimensionless temperatures, $\dt$ the dimensionless time, and $\V^{(i)}$ the dimensionless voltage drops across grain $i$. We also write the conductivities, Eqs.~(\ref{eqs.conduct}), as
\begin{eqnarray}
\label{dimensionless}
\sigma^{(i)}_\inel &=& \frac{2e^2(k_B T_1)^2}{ a\hbar E_c^2} \, \, \tilde\sigma^{(i)}_\inel,\nn \\
\kappa^{(i)}_\inel &=& \frac{\gamma_e (k_B T_1)^3}{a\hbar E_c^2} \, \, \tilde\kappa^{(i)}_\inel, \hspace{0.3cm}
\kappa^{(i)}_\ph = \frac{\gamma_\ph (k_B T_1)^3}{\hbar l_\ph (\hbar \Theta_D)^2} \, \, \tilde\kappa^{(i)}_\ph ,
\end{eqnarray}
where the 'tilde' expressions are dimensionless. The dimensionless current is then $\iota^{(i)}=\tilde\sigma^{(i)}_\inel\V^{(i)}=\frac{\hbar a^2 E_c^2}{2 e (k_B T_1)^3}j^{(i)}$ and we define the dimensionless thermal conductivity as
\begin{equation}\label{eq.dkappa}
\tilde\kappa^{(i)}=\gamma_e\alpha\tilde\kappa^{(i)}_\inel+\beta\tilde\kappa^{(i)}_\ph\,,
\end{equation}
where we introduced the dimensionless parameters\footnote{In typical situations the parameter $\beta$ is much larger than $\alpha$.}
\begin{equation}
\label{alpha}
\alpha = 3 \left(\frac{k_B T_1}{E_c}\right)^2, \hspace{0.3cm}
\beta = 3 \gamma_\ph  \frac{a}{l_\ph } \left(\frac{T_1}{\Theta_D}\right)^2\,.
\end{equation}
Substituting Eqs.~(\ref{dimensionless}) back into Eq.~(\ref{eq.model2}) we obtain
the following dimensionless heat equation
\begin{equation}
\label{eq.Tg2}
\tilde\nabla\left(\tilde\kappa^{(i)}\tilde\nabla \T^{(i)}_g\right) =\T_g^{(i)}\frac{\partial\T_g^{(i)}}{\partial\tau}-2\alpha\frac{\left(\iota^{(i)}\right)^2}{ \tilde\sigma^{(i)}_\inel}+\PSd^{(i)},
\end{equation}
where $\tilde\nabla$ is the dimensionless gradient and the dimensionless electron-phonon coupling term to the substrate is $\PSd^{(i)}=\tilde\Sigma_s\left[\left(\T^{(i)}_g\right)^5 - \T_s^5\right]$ with
$\tilde\Sigma_s=\frac{3\hbar a^3T_1^3}{k_B^2}\Sigma_s$, Ref.~(\onlinecite{RMP06}).
Equation~(\ref{eq.Tg2}) is the main equation which is considered in the following.

\subsection{Discrete kinetic heat equation}

Here we present the explicit expression of the kinetic heat Eq.~(\ref{eq.Tg2}) for the discrete chain of grains.
First, we derive an expression for the dimensionless voltage drop on grain $(i)$, which follows from $\V^{(i)}=\iota^{(i)}/\tilde\sigma^{(i)}_\inel=\Upsilon^{(i)}/[(\T_g^{(i)})^2+\V^{(i)})^2]$ with $\Upsilon^{(i)}=\iota^{(i)}/(g_t^{(i)})^2$, leading to
\begin{equation}\label{eq.Vi}
\V^{(i)}=18^{-1/3} \Xi^{(i)}-(2/3)^{1/3}(\T_g^{(i)})^2/\Xi^{(i)},
\end{equation}
with $\Xi^{(i)}=\left(9 \Upsilon^{(i)}+\sqrt{81 \left(\Upsilon^{(i)}\right)^2+12\left(\T_g^{(i)}\right)^6}\right)^{1/3}$ . The total voltage drop over the chain is then $\V_{\mathrm{tot}}=\sum_i\V^{(i)}$.
The l.h.s of Eq.~(\ref{eq.Tg2}) is explicitly given by (written as central differences; cf. also Eq. (23) of Ref.~[\onlinecite{RMP06}])
\begin{eqnarray}
\tilde\nabla\left(\tilde\kappa^{(i)}\tilde\nabla \T^{(i)}_g\right)&=&\left[\tilde\nabla\tilde\kappa^{(i)}\frac{\T^{(i+1)}_g-\T^{(i-1)}_g}{2}\right.\nn\\
&+&\left.\tilde\kappa^{(i)}\left(\T^{(i+1)}_g+\T^{(i-1)}_g-2\T^{(i)}_g\right)\right]\,.\label{eq.hflux}
\end{eqnarray}
Using Eqs.~(\ref{eq.Vi}) and (\ref{eq.dkappa}) we obtain for the gradient of thermal conductivity $\tilde\nabla\tilde\kappa^{(i)}$ the following expression
\begin{equation}\label{eq.tkappa}
\tilde\nabla\tilde\kappa^{(i)}\equiv\frac{\T^{(i+1)}_g-\T^{(i-1)}_g}{2}\left[\gamma_e\alpha\iota^{(i)}\frac{\partial [\T_g^{(i)}/\V^{(i)}]}{\partial \T_g^{(i)}}+3\beta\left(\T_g^{(i)}\right)^2\right]\,.
\end{equation}
Here we took into account the fact that the tunneling conductance $g_t^{(i)}$ is an averaged quantity for a particular grain $(i)$, with
\begin{equation}
\label{11}
\frac{\partial [\T_g^{(i)}/\V^{(i)}]}{\partial \T_g^{(i)}}=\frac{3\left(\frac{3}{2}\right)^{1/3}\Upsilon^{(i)}\left[12^{1/3}\left(\T_g^{(i)}\right)^2
+\left(\Xi^{(i)}\right)^2\right]}{\Xi^{(i)}\left(\V^{(i)}\right)^2\sqrt{81 \left(\Upsilon^{(i)}\right)^2+12\left(\T_g^{(i)}\right)^6}}\,.
\end{equation}
where $\V^{(i)}$ is given by Eq.~(\ref{eq.Vi}). Using Eqs.~(\ref{eq.hflux})-(\ref{11}) we solve the kinetic heat Eq.~(\ref{eq.Tg2}) numerically for different situations.
In the simulations we keep the lead temperatures $\T_1=\T_g^{(0)}=1$ and $\T_2=\T_g^{(N+1)}$ fixed and integrate (\ref{eq.Tg2}) until a steady state is reached. For the initial condition of the grain temperatures we choose a linear interpolation from $\T_1$ to $\T_2$.

\subsection{Kinetic heat equation: Stationary case}

\begin{figure}[b]
\includegraphics[width=0.95\columnwidth]{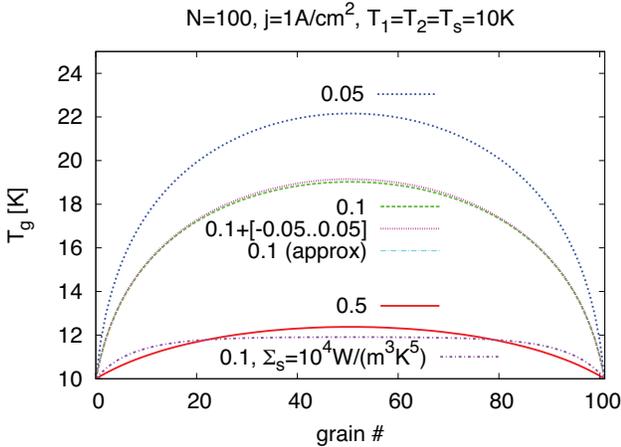}
\caption{(Color online) Stationary heat profiles (grain temperature $T_g$ vs. grain index) for the symmetric case where both lead temperatures coincide $T_1 = T_2 = 10$K and for different tunneling conductances $g_t=0.5,0.1,0.05$. The graph for $g_t=0.1$ is almost identical with the profile for random $g_t^{(i)}$ with mean $0.1$ and variance $0.05$ (dotted, magenta) and the approximation for $g_t=0.1$ (dash-dot, cyan). A curve with enabled coupling to the substrate (see definition below Eq.~(\ref{eq.Tg2})) is shown as well for $g_t=0.1$ and $\Sigma_s=10^4$W/(m$^3$K$^5$) at a substrate temperature of $T_s=10$K. The current density is fixed at $j=1$A/cm$^2$ and length $N=100$.}
\label{fig.Tg_station}
\end{figure}

\begin{figure}[tbh]
\includegraphics[width=0.95\columnwidth]{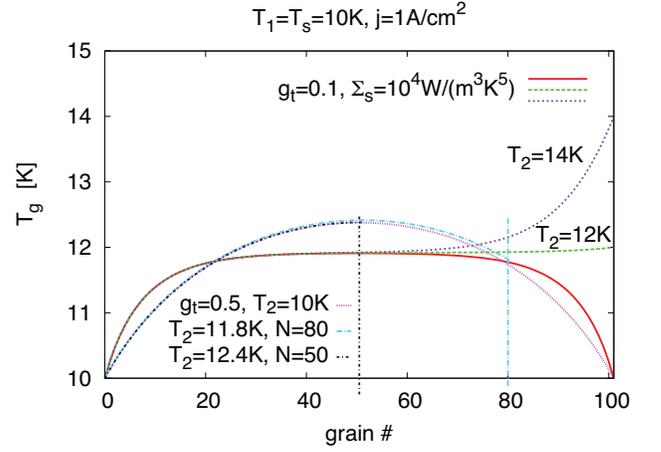}
\caption{(Color online) Stationary heat profiles (grain temperature $T_g$ vs. grain index) for $T_1=T_s=10$K for asymmetric case with different right lead temperature $T_2$, in particular $T_2 > T_1$. The curves for tunneling conductance $g_t=0.5$ for different chain lengths without substrate coupling ($\PSd = 0$) are chosen in such a way that the right-lead temperature $T_2$ lies on the symmetric profile with $T_2=10$K (number of grains $N=100$), i.e. $T_2=12.4$K for $N=50$ and $T_2=11.8$K for $N=80$. For the latter two curves the end is marked by a vertical line since all three curve lie on top of each other.
The second set of three curves is plotted for $g_t=0.1$ and electron-phonon coupling constant $\Sigma_s=10^4$W/(m$^3$K$^5$), for three different right lead temperatures $T_2=10,12,14$K, written next to these curves (all for $N=100$). The current density is fixed at $j=1$A/cm$^2$ for all graphs.}
\label{fig.Tg_T2}
\end{figure}

Before solving kinetic heat Eq.~(\ref{eq.Tg2}) for the dimensionless temperatures $\T^{(i)}_g(\tau)$ vs. time $\tau$ exactly, we discuss its stationary solution
corresponding to large times, $\tau \gg \tau^*$, where $\tau^*$ is some characteristic time scale which we will discuss in section~\ref{sec.dyn}. We also consider only the case $\PSd^{(i)}=0$  (no coupling between the array and the substrate) in this section.
In other words we first estimate how pronounced the effect of inelastic cotunneling on grain heating is.
For large times the grain temperature reaches some constant steady state value,
\begin{equation}\label{eq.Tgs}
\lim_{\tau\to\infty}\T^{(i)}_g(\tau)  \equiv \T^{(i)*}_g\,,
\end{equation}
i.e. $\partial \T^{(i)*}_g/\partial \tau=0$. In this section we consider only the stationary case and suppress the superscript $^*$ in the following.
Using Eq.~(\ref{eq.Tg2}) we obtain
\begin{equation}
\label{eq.station}
\tilde\nabla\left(\tilde\kappa^{(i)}\tilde\nabla \T^{(i)}_g\right)=-2\alpha\frac{\left(\iota^{(i)}\right)^2}{ \tilde\sigma^{(i)}_\inel}\,.
\end{equation}

In the limiting cases of low voltage $(\V^{(i)} \ll \T^{(i)}_g)$ or low temperature $(\T^{(i)}_g \ll \V^{(i)})$ one can find the analytical solution of Eq.~(\ref{eq.station}) for the dimensionless grain temperature $\T^{(i)}_g$.
However, due to the cumbersome expression for the discrete heat-flux, Eq.~(\ref{eq.hflux}) we go over to the straight forward continuum approximation, i.e. the grain index $i$ goes over in coordinate $x$ and the discrete gradients become continuum ones.

\subsubsection{Low voltages}

We first consider the stationary solution, Eq.~(\ref{eq.station}), in the limit of small voltages, $\V \ll \T_g$.
Using Eqs.~(\ref{eqs.conduct}), (\ref{dimensionless}) and (\ref{eq.station}) we obtain
\begin{equation}
\label{smallV}
\tilde\nabla [\T_g^3 \tilde\nabla \T_g]=-\frac {\Upsilon_L^2}{\T_g^2},
\end{equation}
where we introduce the notation $\Upsilon_L^2 \equiv \frac{ 2\left(\iota\right)^2}{g_t^2 ( g_t^2\gamma_e + \beta/\alpha)}$.
Equation~(\ref{smallV}) can be integrated using the new variable $z(\T_g) = \T_g^3 \tilde\nabla \T_g$. Then $\tilde\nabla z = z' \tilde \nabla \T_g=z'z/\T_g^3$, where $z' = dz/d\T_g$. In terms of variable $z$, Eq.~(\ref{smallV}) has the form
\begin{equation}
\label{zz1}
z'z=-\Upsilon_L^2{\T_g},
\end{equation}
leading to the solution $z=\Upsilon_L\sqrt{ \T_m^2-\T_g^2}$, with $\T_m$ being the integration constant, physically meaning $\max_x\T_g(x)$  which has the coordinate $x_m$, i.e. $\T_m=\T_g(x_m)$.
Using this solution we obtain the following differential equation for the dimensionless temperature
\begin{equation}
\frac{\T_g^3\nabla \T_g}{\sqrt{ \T_m^2-\T_g^2}}=\Upsilon_L.
\end{equation}
Integration of this equation $\int_{\T_g}^{\T_m}\frac{\T_g^3\,d\T_g}{\sqrt{ \T_m^2-\T_g^2}}=(x_m-x){\Upsilon_L}$ leads to the following solution
\begin{equation}
\label{solution1}
\frac13\sqrt{ \T_m^2-\T_g^2}( 2\T_m^2+\T_g^2)=(x_m-x){\Upsilon_L}.
\end{equation}
Introducing the new variable $y = \T_g^2/\T_m^2$, Eq.~(\ref{solution1}) can be written as follows
\begin{equation}
\label{cubic}
    y^3 + 3 y^2 = 4 - (x-x_m)^2 (3\Upsilon_L/\T_m^3)^2 \equiv s+2.
\end{equation}
The positive root of Eq.~(\ref{cubic})  reads
\begin{equation}
\label{solution2}
y_3 = C_{1/3}\left(s\right)-1\,, 
\end{equation}
where we introduced Chebyshev's cubic root function, see Appendix~\ref{ap1}
\begin{gather}
C_{1/3}(s)=2\cosh\left[\frac13\acosh(s/2)\right].
\end{gather}
Using Eq.~(\ref{solution2}) for the dimensionless grain temperature we finally obtain
\begin{equation}
\label{result1}
\frac{\T_g (x)}{\T_m} = \sqrt{C_{1/3}\left(s\right)-1},
\end{equation}
where parameter $\Upsilon_L$ is defined below Eq.~(\ref{smallV}). For the symmetric case (both lead temperatures are equal,
$\T_1 = \T_2$) Eq.~(\ref{result1}) is plotted in Fig.~\ref{fig.Tg_station}. One can see that the maximum temperature $\T_m$
is reached in the middle of the chain.

\subsubsection{High voltages}

Here we discuss the stationary solution,  Eq.~(\ref{eq.station}), in the limit of small temperatures, $ \T_g \ll \V$.  Using Eqs.~(\ref{dimensionless}) and (\ref{eq.station}) we obtain the following equation
\begin{equation}
\label{highV}
\tilde\nabla [\T_g \tilde\nabla \T_g] = - \Upsilon_H^2,
\end{equation}
where we introduce the notation $\Upsilon_H^2 \equiv \frac{2 \iota^2}{\gamma_e g_t^4 \V^4}$. Equation~(\ref{highV}) can be integrated using the new variable, $z(\T_g)=\T_g \tilde\nabla \T_g$. Then $\tilde\nabla z =z' \tilde\nabla \T_g=z'z/\T_g$, where $z' = dz/d\T_g$. In terms of variable $z$, Eq.~(\ref{highV}) has the form
\begin{equation}
\label{zz}
 z'z = - \Upsilon_H^2{\T_g},
\end{equation}
leading to the solution $z=\Upsilon_H\sqrt{ \T_m^2-\T_g^2}$. Equation~(\ref{zz}) is similar to Eq.~(\ref{zz1}) obtained in the large temperature limit. The solution of Eq.~(\ref{zz}) reads
\begin{equation}
\label{result2}
\frac{\T_g(x)}{\T_m} = \sqrt{C_{1/3}\left(s\right)-1}. 
\end{equation}
with $s=2 - (x-x_m)^2 (3\Upsilon_H/\T_m^3)^2$.

\subsection{Temperature-voltage dependence}

\begin{figure}[htb]
\includegraphics[width=0.8\columnwidth]{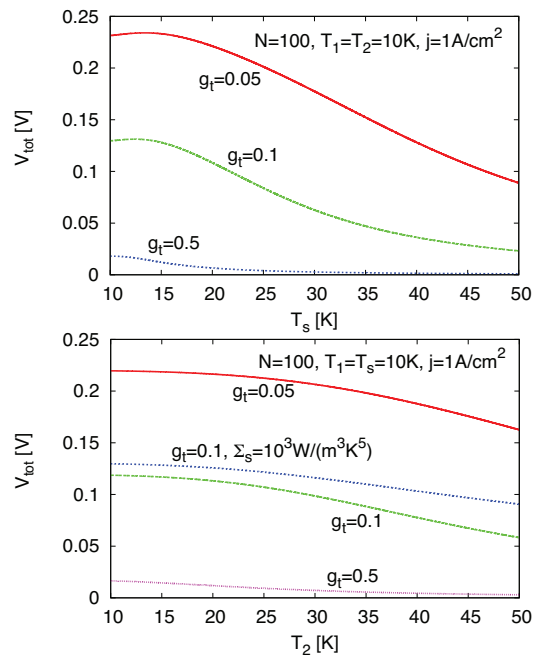}
\caption{(Color online) Voltage dependence of nano dot chain  $V_{\mathrm{tot}}$ vs substrate temperature $T_s$ (top) and right lead temperature $T_2$ (bottom) for different tunneling conductances $g_t$. All plots are for applied current density $j=1$A/cm$^2$, left lead temperature $T_1=10$K, and chain length $N=100$. In the upper panel the electron-phonon coupling constant is $\Sigma_s=10^3$W/(m$^3$K$^5$). In the lower plot the electron-phonon coupling is switched off, apart from the dotted blue curve marked with $g_t=0.1, \Sigma_s=10^3$W/(m$^3$K$^5$) which is coupled to a substrate at $T_s=10$K. }
\label{fig.VT}
\end{figure}

As mentioned in the introduction the chains of nano dots can be used as highly sensitive thermometers from cryogenic up to room temperature. In order to demonstrate this, we studied the total voltage drop, $V_{\mathrm{tot}}$, on the chain for different substrate $T_s$ or lead $T_2$ temperatures. The numerical results for $V_{\mathrm{tot}}$ are shown in Fig.~\ref{fig.VT} for $T_s$ and $T_2$ in the range from $10$K to $50$K.
As one can see, the chains are particularly sensitive if the substrate (sample) is slightly above the lead temperature. It is clear that the sensitivity is reduced if only one lead is used as temperature sensor.
In the case of the substrate sensor, the sensitivity range can be tuned by the tunneling conductance of  the chain.
Another important issue for applications as temperature sensors is the response time of the device. In the case of the chain of grains, this response time can be reduced by (i) increasing the current density, $j$, (ii) having a better electron-phonon coupling to the sample, i.e. larger $\Sigma_s$, (iii) or smaller tunneling conductances, $g_t$; see next section.
Timescales of several nanoseconds are typical for not too long chains. The length of the chain can be adjusted for best sensitivity and fastest response time.

\subsection{Kinetic heat equation: Dynamic case}\label{sec.dyn}

\begin{figure}[tbh]
\includegraphics[width=0.98\columnwidth]{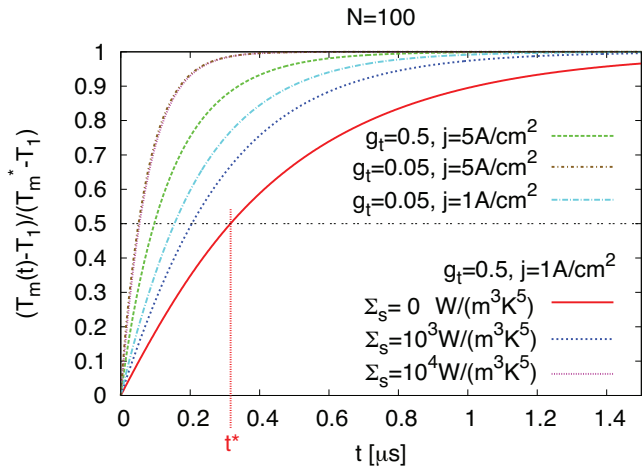}
\caption{(Color online) Plot of the rescaled maximum grain temperature $T_m(t)$ vs. time $t$. The typical timescale $t^*$ to reach the steady state value $T_m^*$ is shown for tunneling conductance $g_t=0.5$ and current density $j=1$A/cm$^2$ (solid, red curve). For comparison also the graphs for two different substrate coupling are shown: for larger coupling the typical timescale $t^*$ is reduced. The plot also shows the time evolution of $T_m(t)$ for different $g_t$ and $j$ (see legend) for zero electron-phonon coupling (no substrate): for smaller $g_t$ or larger $j$, the timescale $t^*$ can be decreased as well.}
\label{fig.Tg_t}
\end{figure}

\begin{figure}[tbh]
\includegraphics[width=0.7\columnwidth]{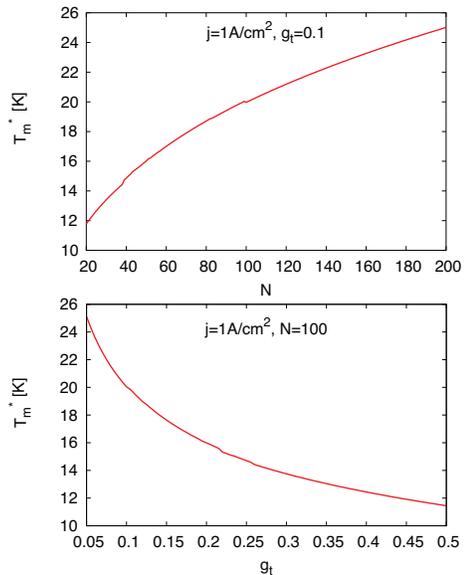}
\caption{(Color online) Upper panel: Solution of Eq.~(\ref{result11}) for maximum temperature $\T^*_m$ vs. lengths $N$ of nano dot chain for a fixed tunneling conductance $g_t=0.1$. Low panel:  $\T^*_m$ vs $g_t$ for a fixed length $N=100$. Both graphs are plotted for constant current density $j=1$A/cm$^2$.}
\label{fig.Tm}
\end{figure}

In this section we estimate the characteristic dimensionless timescale $\dt^*$ [see Eq.~(\ref{dimensionless1})] associated with the heating of nano dot chain due to inelastic cotunneling. We define $\tau^*$ as the typical timescale to reach the maximum temperature $\T^*_m$ in the chain,
$\T^*_m  =\max_i \T_g^{(i)*}= \lim_{\dt\rightarrow \infty} \T_{m}(\dt)$. The time $\dt^*$ can be easily estimated for low voltages $\V \ll \T$. Using the fact that for symmetric case (both leads have the same temperature, $\T_1 = \T_2$) the maximum temperature $\T_m$ is located in the middle of the chain ($i=N/2$), meaning that $\tilde\nabla \T_m = 0$, and
using Eq.~(\ref{eq.Tg2}) in the absence of a substrate, $\PSd^{(i)} = 0$, we obtain the following equation
\begin{equation}
\label{time1}
\T_m \frac{\partial\T_m}{\partial\tau} = \frac{2\alpha \,  \iota^2}{ \tilde\sigma^{(N/2)}_\inel}.
\end{equation}
Integrating Eq.~(\ref{time1}) for a given fixed current $\iota$ and using Eq.~(\ref{dimensionless}) for the dimensionless conductivity $\tilde\sigma^{(N/2)}_\inel$ we obtain the following result
\begin{equation}
\label{dts}
\tau^* = \frac{g_t^2}{8 \alpha \iota^2} (\T_m^*)^4.
\end{equation}
Here the maximum temperature $\T_m^*$ is implicitly defined by the following equation [see also Eq.~(\ref{result1})]
\begin{eqnarray}
1& =& \T_m^* \sqrt{C_{1/3}(s) - 1 }\,, \,\,\text{with}\label{result11}\\
s&=&2 - \frac{18\,  (N/2)^2 \, \iota^2 }{g_t^2(g_t^2\gamma_e + \frac{\beta}{\alpha} )(\T_m^*)^6}\,.\nn
\end{eqnarray}
The solution of Eq.~(\ref{result11}), $\T_m^*$ vs number of grains, $N$, and $\T_m^*$ vs. tunneling conductance, $g_t$, is shown in Fig.~\ref{fig.Tm}.

\section{Over-heating hysteresis}

\begin{figure}[tbh]
\includegraphics[width=0.9\columnwidth]{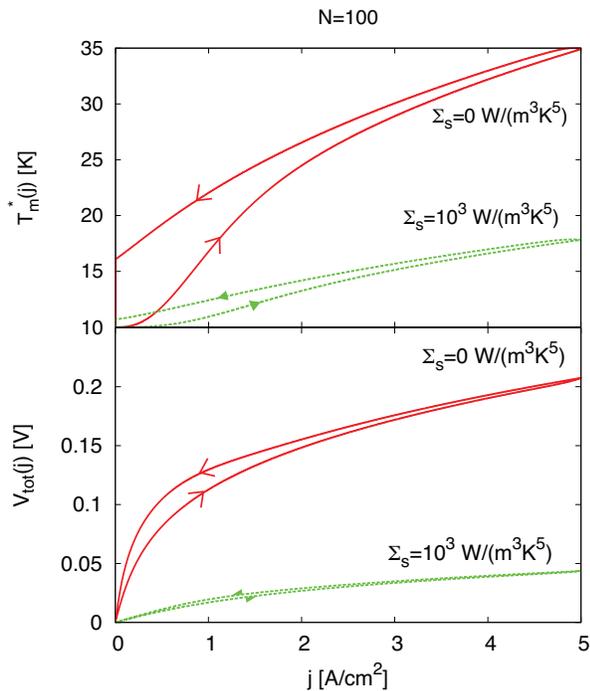}
\caption{(Color online) Hysteretic behavior of the maximum temperature $T_m^*$ (upper panel) and the total voltage trop $V_{\mathrm{tot}}$ (lower panel) vs current density $j$. The current density $j$ was first increased from $0$A/cm$^2$ to  $5$A/cm$^2$ and then decreased back to $0$A/cm$^2$. The arrows show the direction of the hysteresis. The red curves correspond to the case with no coupling to the substrate. The green curves are shown in the presence of the
substrate, with electron-phon coupling constant $\Sigma_s=10^3$W/(m$^3$K$^5$). The tunneling conductance for all curves was fixed, $g_t=0.1$.}
\label{fig.hysteresis}
\end{figure}

If the rate of the energy supply from the external bias to the charge carriers exceeds the rate of energy losses to the environment, the phenomenon of \textit{overheating} appears and the energy distribution function of the current carriers noticeably deviates from the equilibrium distribution
function.~\cite{heatingReviewA,heatingReviewB} One of the characteristic manifestations of the overheating effect is the onset of the ``falling'' region of the $I$-$V$ curve where the differential conductivity  $G={\partial I}/{\partial V}<0$. The corresponding $I(V)$ characteristics is referred to as that of the $S$-type if the current is the multi-valued function of the voltage, and the $I$-$V$ curve of the $N$-type corresponding to the case where the current is nonmonotonic but still remains a single valued function of the voltage. The phenomenon of overheating has been a subject of the incremental interest and extensive studies during the decades, see review of Volkov and Kogan,~\cite{heatingReviewA} and the impressive progress in understanding of the underlying mechanisms was achieved. Recently it was suggested~\cite{Ovadia} that the mechanism of heating instability described in detail in Ref.~\onlinecite{heatingReviewA} combined with the expression for the dissipated power calculated for strongly disordered conductors~\cite{Schmid,Reizer-JETP,Reizer1989,Sergeev} (see Ref.~\onlinecite{RMP06} for review)
can explain switching $I$-$V$ characteristics in InO and TiN samples.

Using Eq.~(\ref{eq.Tg2}) we simulate the hysterestic overheating behavior in chains of nano-grain arrays. We first increase the current from $j=0$A/cm$^2$ to $5$A/cm$^2$ and then decrease it back to zero. The results are shown in Fig.~\ref{fig.hysteresis}. One can see a pronounced hysteretic behavior of both the total voltage drop $V_{\mathrm{tot}}$ and the maximum chain temperature $T_m^*$. It is important to remark that this hysteretic behavior disappears for shorter chains.

\section{Discussions}

In all simulations throughout this work, we use typical experimental parameters to estimate the heating effects on a chain of weakly coupled grains and presented all equations in either SI or dimensionless units. In particular we used the following physical parameters: the grain size $a=10$nm, Debye temperature $\Theta_D\approx 450$K, Coulomb energy $E_c\approx 1600$K (with relative static
permittivity $\kappa=1$), mean energy level spacing  $\delta=1$K, phonon mean free path $l_\ph=a$,  and tunneling conductance $g_t\in [0.05;0.5]$. Furthermore we simulated typical chains of $N=100$ grains (i.e. of a few $\mu$m  total length).

For the electron-phonon coupling to the substrate we used values of $10^3$ or $10^4$W/(m$^3$K$^5$) since we want to describe only the weak coupling of the chain to a different substrate.
For comparison, typical values for bulk materials are much larger (on the order $10^9$W/(m$^3$K$^5$), see e.g.~[\onlinecite{RMP06}]).

As for the analytical approximations in the low- and high-voltage regions (without substrate coupling) the particular form of the differential and later algebraic equations, allowed us to derive a closed solution using
Chebyshev's cube root.

As a first step we calculated the stationary heat profile of the chain under different conditions. It is important to remark that the low-voltage approximation fits very well the full simulation and randomness in the tunneling conductance has only minor effects on the final profile (provided the chain is not ''broken'' by a very small tunneling conductance.)
Furthermore we estimated the characteristic heating timescale $t^*$ and maximum temperature $T_m^*$ in Eq.~(\ref{dts}) using the analytical solution and obtained as a result $t^* \approx 0.1 \mu$s and $T_m^* \approx 20$K.

These estimates can be directly compared with our simulations of Eq.~(\ref{eq.Tg2}) for rescaled maximum grain temperature $T_m(t)$ vs. time $t$. The outcome of these simulations is shown in Fig.~\ref{fig.Tg_t}. One can see that the typical timescale $t^*$ to reach the steady state value $T_m^*$ is of order of $0.1 \mu$s. For completeness in Fig.~\ref{fig.Tg_t} we also presented several different curves $T_m(t)$ vs. time $t$ for different tunneling conductances $g_t$ and current densities $j$. In addition, we also presented two curves in the presence of the substrate: for larger coupling the typical timescale $t^*$ is reduced. Based on the above there is a good agreement between our estimates and simulations.
At this point we note that in general the kinetic heat Eq.~(\ref{eq.Tg2}) has an additional term proportional to the thermoelectric coefficient. Recently, thermoelectric properties of weakly coupled granular arrays were studied~\cite{glatz+prb09b,Glatz09}. In particular, in Ref.~[\onlinecite{Glatz09}] we estimated the thermoelectric and Seebeck coefficients. It was found that the thermoelectric coefficient has an additional small factor proportional to $T/E_F \ll 1$ compared to the electric and thermal conductivities, where $E_F$ is the Fermi energy, i.e., for the temperatures under consideration this additional term in Eq.~(\ref{eq.Tg2}) would be about three orders of magnitude smaller than the others.

In conclusion, we have studied heating effects in one-dimensional chains of quantum dots due to inelastic electron cotunneling using a kinetic heat equation. In the low and high voltage limits we solved the stationary heat equation analytically. We demonstrated the possible application of chains of nanograins as highly sensitive thermometers and estimated the typical timescale to reach a steady heat profile in these chains.
Finally, we showed the over-heating hysteresis in the large-current or voltage regimes.
The influence of the electron-phonon coupling of the chain to a substrate was considered numerically and has pronounced effects on the stationary heat profile, the typical timescales to reach it, and the thermometric and hysteretic behavior.

\acknowledgements
A.~G., N.~C., and V.~V. were supported by the U.S. Department of Energy Office of Science under the Contract No. DE-AC02-06CH11357.
I.~B. was supported by an award from Research Corporation for Science Advancement.


\appendix

\section{Cube roots\label{ap1}}
\begin{figure}
  \includegraphics[width=0.4\textwidth]{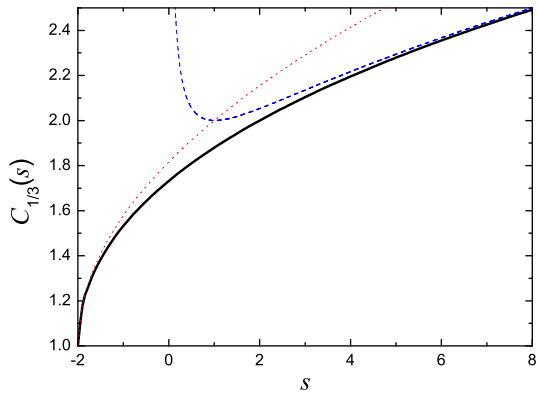}\\
  \caption{(Color online) The graph of $C_{1/3}(s)$ (solid line). The graphs with the dashed and dotted lines show the behavior of the asymptotical approximations, see Eq.~\eqref{eq:C_asympt}.}\label{fig:C13}
\end{figure}

Here we discuss the positive roots of the cubic equation
\begin{equation}
\label{cubic11}
    y^3 + 3 y^2 = s+2.
\end{equation}
The roots of the cubic equation can be always found analytically however the expressions are quite cumbersome. In this paper [see Eq.~(\ref{cubic})] we can remarkably find a simple expression for the root because Eq.~\eqref{cubic11} reduces to the problem of finding roots of the Chebyshev polynomial.  The transformation $y=\tilde y-1$ reduces the left side of Eq.~\eqref{cubic11} to the third Chebyshev polynomial, $C_3(y)=\tilde y^3 -3 \tilde y$, and the right side to $s$. Then the resulting equation has only one positive root known as the Chebyshev cube root:~\cite{Chebyshev} $\tilde y=C_{1/3}(s)$, where
\begin{gather}
     C_{1/3}(s)=2\cosh\left[\frac13\acosh(s/2)\right].
\end{gather}
Here $\acosh(s/2)=\ln[(s^2+\sqrt{s^2-4})/2]$ and we use the branch of the logarithm which is real on the positive real line and the branch of the square root which is positive on the real axis. Asymptotic behavior of the  Chebyshev cube root is the following:
\begin{gather}\label{eq:C_asympt}
    C_{1/3}(s)\approx \left\{
                        \begin{array}{ll}
                          1+\sqrt{\frac{s+2}3}, & \hbox{if $0<s+2\ll 1$;} \\
                          2\cosh\left(\frac {\ln s}3\right), & \hbox{$s\gg 1$.}
                        \end{array}
                      \right.
\end{gather}
The graph of $C_{1/3}$ is shown in Fig.~\ref{fig:C13}

\end {document}